\documentstyle[11pt,aaspp4]{article}

\begin{document}

\title{Hubble Space Telescope Observations of the CfA Seyfert 2s: The 
Fueling of Active Galactic Nuclei\altaffilmark{1}}

\author{Paul Martini \& Richard W. Pogge}

\affil{Department of Astronomy, 174 W. 18th Ave., Ohio State University, \\
Columbus, OH 43210 \\
martini,pogge@astronomy.ohio-state.edu}

\altaffiltext{1}{Based on observations with the
NASA/ESA {\it Hubble Space Telescope} obtained at the the Space Telescope
Science Institute, which is operated by the Association of Universities for
Research in Astronomy, Incorporated, under NASA contract NAS5-26555.}

\begin{abstract}

We present an investigation of possible fueling mechanisms
operating in the inner kiloparsec of Seyfert galaxies.  We analyze
visible and near-infrared {\it Hubble Space Telescope} images of 24
Seyfert 2s from the CfA Redshift Survey sample.
In particular, we are searching for the morphological signatures of
dynamical processes reponsible for transporting gas from kiloparsec
scales into the nucleus.  The circumnuclear regions are very rich in
gas and dust, often taking the form of nuclear spiral dust lanes on
scales of a few hundred parsecs.  While these nuclear spirals are
found in 20 of our 24 Seyferts, we find only 5 nuclear bars among the entire
sample, strongly reinforcing the conclusions of other investigators that
nuclear bars are not the primary means of transporting this material
into the nucleus.  An estimate of the gas density in the nuclear
spirals based on extinction measurements suggests that the nuclear
spiral dust lanes are probably shocks in nuclear gas disks that are
not strongly self-gravitating.  Since shocks can dissipate energy and
angular momentum, these spiral dust lanes may be the channels by which
gas from the host galaxy disks is being fed into the central engines.

\end{abstract}


\clearpage 

\section{Introduction}

An important unsolved problem in the study of active galactic nuclei (AGN)
is how the central massive black hole is fueled.  In particular, if the
primary fuel source is interstellar gas and dust in the host galaxy, how is
that material transported from kiloparsec scales into the central parsecs
of the galaxy and onto the supermassive black hole? In order to fuel the AGN,
this material must somehow lose nearly all of its angular momentum.
Estimates of the mass accretion rate to power AGN range
from more than $1\;M_{\sun}$ per year for the most luminous quasars to $\sim
0.01\;M_{\sun}$ per year for the Seyfert galaxies discussed here.  Thus, over
an AGN's lifetime (10$^{7-8}$ years?) a significant amount of material must
be transported inwards to feed the central black hole.

A number of dynamical mechanisms have been proposed that could remove
angular momentum from the host galaxy's gas and thus supply the fueling
rates required to match the observed luminosities of AGN.  These include
tidal forces caused by galaxy-galaxy interactions (Toomre \& Toomre 1972)
and stellar bars within galaxies (Schwartz 1981).

Numerical simulations of galaxy interactions (Hernquist 1989; Barnes \&
Hernquist 1991; Mihos, Richstone, \& Bothun 1992; Bekki \& Noguchi 1994;
Bekki 1995) have shown that strong perturbations on a galaxy disk due to a
close encounter with another galaxy can trigger the infall of large amounts of
gas into the central 1 kpc to 10 pc, depending on the simulation resolution,
the type of encounter, and the properties of the simulation.  Further,
relatively minor encounters (``minor mergers'', Hernquist 1989; Mihos \&
Hernquist 1994; Hernquist \& Mihos 1995) can lead to
significant gas infall, so one need not invoke disruptive encounters.
Observationally, there is ample evidence of past or on-going tidal
encounters and companions among nearby luminous QSOs (Hutchings \& Neff
1988, 1992, 1997; Bahcall et al. 1995a,b, 1997), but at most only a mild
statistical excess of interactions or close companions is seen among the
lower-luminosity Seyfert galaxies (Adams 1977; Petrosian 1982; Dahari 1984;
Keel et al. 1985; Keel 1996; DeRobertis, Hayhoe, \& Yee 1998).

In barred galaxies, the triaxial stellar potential leads to a family of
orbits for the stars and gas that can transport material into the inner
100$-$1000 pc of the galaxy.  Models of orbits using 2-D and 3-D
hydrodynamic simulations of gas and stars in barred potentials
(Athanassoula 1992; Friedli \& Benz 1993; Piner, Stone, \& Teuben 1995)
have shown the formation of a shock front at the leading edge of a bar.
Material builds up in this shock and falls into the nuclear region,
generally forming a nuclear ring with a diameter approximately equal to the
bar's minor axis ($>$100$-$1000 pc).  Observations of apparent shock fronts
at the leading edges of bars and velocity distributions in barred galaxies
are in general agreement with inflow models (Quillen et al. 1995; Regan,
Vogel, \& Teuben 1997), and provide evidence that the dynamical behavior
predicted by the models is realistic.

However, the typical semiminor axis of a bar is still $\sim$100$-$1000
pc, and thus while bars might plausibly transport material reasonably close
to the center, an additional mechanism is needed to move the material into
the nucleus proper.  A mechanism proposed to complete the transport of
material into the nucleus is the `bars within bars' scenario (Shlosman,
Frank, \& Begelman 1989; Maciejewski \& Sparke 1997; Erwin \& Sparke 1998).
In this model, a large-scale bar transports material into a kiloparsec-scale
disk.  This gaseous disk then becomes unstable to bar formation, creating a
miniature `nuclear' bar that transports the gas to within approximately 10 pc
of the galactic nucleus, which is approximately where the central supermassive
black hole's potential can take over.  While large-scale bars in galaxies are
readily observable due to their large angular size, small-scale, nuclear
bars (Shaw et al. 1995) are a factor of 5 to 10 smaller and thus are more
difficult to resolve in distant galaxies.  Because of their smaller
physical size, nuclear bars were initially discovered in relatively nearby
galaxies (de Vaucouleurs 1975; Kormendy 1979; Buta 1986a,b).

Observationally, investigations of the relative fraction of bars in AGN as
compared to non-active galaxies have failed to turn up a significant excess
of bars among spirals harboring AGN (Kotilainen et al. 1992; McLeod \& Rieke
1995; Alonso-Herrero, Ward, \& Kotilainen 1996; Mulchaey \& Regan 1997).
Similarly, searches for nuclear bars using the {\it Hubble Space Telescope}
({\it HST}) have failed to find them in numbers sufficient to salvage the bar
fueling picture (Regan \& Mulchaey 1999).  We shall reach a similar
conclusion in this paper.  At most, these observations have shown that nuclear
bars are present in only a minority of Seyferts.

This is not to say that bars and interactions are not viable fueling mechanisms
as clearly they {\it are} operating in a number of Seyferts.  The point is that
neither is the dominant mechanism, nor necessarily responsible for fueling the
AGN in all cases.  What is responsible for fueling an AGN in the majority of
cases where there is no evidence for either interactions or bars?  One way
of addressing this question is to assay the amount of potential fuel in the
form of interstellar dust and gas in the inner kiloparsec of nearby Seyfert
galaxies.  Indeed, while Regan \& Mulchaey (1999) failed to find a
preponderance of nuclear bars in their {\it HST} imaging sample of 12 Seyferts,
they {\it did} find one common morphological structure: spiral arms of dust, of
greater or lesser degrees of organization, in the inner few hundred parsecs.

In this paper we present new, high angular resolution {\it HST} near-infrared
NICMOS images combined with archival WFPC2 images
of the nuclear regions of a representative sample of 24 Seyfert 2s to
search for dynamical and morphological signatures of AGN fueling.  In
section 2 we describe our sample selection, and our observations in section
3.  In section 4 we describe the nuclear morphology derived from direct
flux and color maps to search for nuclear bars and circumnuclear dust
structures.  Section 5 describes the possible role of these structures in
fueling the AGN, and we summarize our results and conclusions in section 6.

\section{Sample Selection}

The history of the study of interactions among Seyfert galaxies serves to
illustrate the main difficulty in drawing conclusions on the nature of AGN from
studies of statistical samples, namely the problem of defining an unbiased
sample and identifying a suitable control group.  As discussed by Keel
(1996), the reexamination of interacting galaxy samples illustrated the
importance of matching host galaxy types and luminosities (Fuentes-Williams
\& Stocke 1988).  In the last decade, better defined samples of AGN,
for example the Center for Astrophysics (CfA) sample (Huchra \& Burg 1992)
and the Revised Shapley-Ames (Sandage \& Tammann 1987) sample of Seyfert
galaxies (Maiolino \& Rieke 1995), have greatly assisted statistical studies of
the AGN population.

The CfA Redshift Survey (Huchra et al. 1982) obtained spectra of a complete
sample of 2399 galaxies down to a limiting photographic magnitude of
$m_{Zw}\le 14.5$.  This relatively unbiased survey avoids many of the
problems of traditional surveys for AGN, particularly those based on
ultraviolet excesses which are biased against reddened AGN and thus
preferentially
detecting Seyfert 1s and Seyferts in host galaxies with active star
formation.  This sample also has the often overlooked advantage of a
uniform set of spectral classifications obtained with high signal-to-noise
ratio that provides comparable detection limits for weak broad-line
components (Osterbrock \& Martel 1993).  This is essential for making
meaningful classifications of Seyfert types 1.8, 1.9, and 2 (Osterbrock
1981) for distinguishing objects with and without obvious broad-line
components.

Studies of the relative frequency of close companions among the CfA
Seyferts have shown that this sample has only a marginal excess of
companions, although a larger fraction of the Seyferts do appear to be
currently undergoing the final stages of a past interaction (DeRobertis,
Hayhoe, \& Yee 1998).  In contrast, this same study emphasizes that a small
but significant number of the CfA Seyferts show no morphological evidence
for recent interactions.
Infrared wavelength imaging also shows no signature of a violent interaction
history in this sample (McLeod \& Rieke 1995), nor does it detect an excess of
nuclear bars.

For this study, we have chosen 24 of the 25 Seyfert 2s (including types 1.8
and 1.9 as classified by Osterbrock \& Martel 1993) in the CfA sample.  We
are concentrating on just the Seyfert 2 galaxies because they in general
have fainter nuclear point sources than Seyfert 1s (Nelson et al. 1996;
Malkan, Gorjian, \& Tam 1998), and thus the circumnuclear environment is
relatively unaffected by the contamination from the complex {\it HST}
point-spread function (PSF).  This is especially true with NICMOS where the
PSF is very complex (MacKenty et al. 1997).

\section{Observations}

Our observations are comprised of archival visible-band and new
near-infrared {\it HST} images.  The visible wavelength WFPC2
images include all of the Seyfert 2 galaxies in the CfA sample with the
exception of NGC\,4388 and Mrk\,461 (neither have been observed to date
with broad-band filters).  Most of the visible images were from a snapshot
survey of AGN (GO-5479, see Malkan, Gorjian, \& Tam 1998) and are single
500s exposures taken through the F606W filter.  The remaining archival
images were taken through the F547M filter from a variety of programs.
Table 1 lists the properties of the sample galaxies and which
of these two visible band filters (hereafter referred to collectively as the
$V$ filter) was used to obtain the $V-$band galaxy image used in this
investigation.  The images were all obtained with the nucleus roughly
centered on the PC1 detector of WFPC2, which has a plate scale of
0\farcs04553~pixel$^{-1}$ and a total effective field of view of 35\farcs6
(Biretta et al. 1996).

We obtained near-infrared $J-$ and $H-$band images of 23 of our Seyfert
sample with the {\it HST} NICMOS Camera 1 during Cycle 7 (Project GO-7867).
The remaining galaxy in our sample, NGC\,1068, was observed by the NICMOS team
as part of their Guaranteed Time Observations.
We chose Camera 1 as its
plate scale (0\farcs043~pixel$^{-1}$) is closest to that of the PC1 camera
($\sim$~0\farcs046~pixel$^{-1}$),
although with a substantially smaller field of view ($\sim$11\arcsec).  The
near match in plate scale between the visible and infrared images is
helpful for matching images from the two cameras and achieving the best
sampling of the resolutions of the two cameras.
Further, the surface brightness of the underlying host galaxies falls off
sufficiently fast that we would have obtained less signal-to-noise in the
outer regions of NICMOS Camera 2 at a cost of coarser pixel sampling
(0\farcs075~pixel$^{-1}$).  Our images were taken through both the F110W
and F160W filters (hereafter $J$ and $H$, respectively), which are each
approximately 2000 \AA\ wide and are centered at 1.1 $\mu$m and 1.6 $\mu$m.

In each filter we acquired images at four dither positions separated by
1\arcsec\,(SPIRAL-DITH pattern) so that the effects of bad pixels and other
detector artifacts could be eliminated.  At each dither position we used an
exposure ramp (STEP128) for a total of 256 seconds per position.  This
allows us to correct for any saturated pixels that might occur if the
nucleus is brighter at near-IR wavelengths than suggested by the archival
WFPC2 imaging (as would be expected if the nuclei are dusty).  The final
shifted and combined frames have a cumulative exposure time of 1024 seconds
per filter.  This was done for each galaxy, resulting in a
homogeneous set of relatively deep near-infrared images of the central
11\arcsec\,of these galaxies.

To calibrate this dataset we had to perform several additional processing
steps beyond the standard {\it HST} data-reduction pipeline. For the WFPC2
images, this consisted of cosmic$-$ray removal and absolute flux
calibration. The flux calibration included transforming our
F547M and F606W to the Johnson/Cousins system by using the
STSDAS SYNPHOT package to convolve a series of galaxy templates with both
Johnson $V$ and the {\it HST} filters.

For the infrared images we shifted and added the four individual exposures
from the CALNICA part of the data reduction pipeline (MacKenty et al. 1997)
after masking out individual bad pixels and detector artifacts. We chose not
to use the CALNICB part of the standard data reduction pipeline as this
task attempts to subtract the background from the final mosaic image. As our
observations were not through filters contaminated by thermal emission,
this step was not necessary. Furthermore, the background subtraction
attempted by the CALNICB pipeline was in fact subtracting the flux level
due to extended emission from these galaxies, a problem that was particularly
acute for our brighter targets that covered the entire array.  We finally
flux-calibrated our data using the transformations derived by Stephens
et al. (2000) to place the near-infrared images on the CIT system. We will
present the full atlas of our observations in a future publication.

\section{Nuclear Morphology}

\subsection{$V-H$ Color Maps}

If there is cold gas flowing into the nuclear ($<10$ pc) regions from the
host galaxy, we should see evidence of this material in the form of
dust lanes extending from large scales into the nuclear regions.
Dust is a useful tracer of this material as it is generally well-mixed with
gas and can be detected by its attenuation of light from background
stars.  Our infrared images primarily trace the stellar distribution in these
galaxies, while the visible wavelength images trace both the stellar
distribution and the dust.  We combined these images to form $V-H$ color maps
of all of the galaxies for which we had visible band {\it HST} data to map
the dust distribution.

Figure 1 shows the $V-H$ color maps for the Seyfert 1.8s and 1.9s in our
sample, while Figure 2 shows the $V-H$ color maps for the Seyfert 2s.
In Figure 3 we show the $J-H$ color maps for Mrk\,461 and NGC\,4388, the two
galaxies for which we do not have visible-band {\it HST} imaging.
These images are 5\arcsec\,on a side, corresponding to projected spatial
sizes between 2.7 kpc and 125 pc in the galaxies for
$H_0 = 75$ km s$^{-1}$ Mpc$^{-1}$.  The 2$-$pixel
resolution element is 0\farcs091, corresponding to projected spatial scales
between 50 pc and 2.3 pc. In Table 1, column 8 we list the physical scale
corresponding to 1\arcsec\,at the distance of each galaxy.  To provide a
visual reference, a bar showing 100~pc projected distance at the galaxy is
drawn in the bottom left corner of each frame.

To exactly match the plate scales of the two images, we rebinned our NICMOS
images to the plate scale of the PC1 chip using a geometric transformation.
Though the resolutions of the two cameras
are different, we did not smooth our $V$ and $H$ images to a common resolution.
As a result our $V-H$ color maps, particularly those galaxies with a bright
nucleus at $H$, show artifacts of the resolution mismatch as small, ``red''
rings at the position of the nucleus (e.g. Mrk\,334, Mrk\,744)
as well as ``red'' spikes from the NICMOS diffraction pattern (e.g. Mrk\,334).
We did not smooth the $V$ and $H$ images to a common resolution as we found
that this substantially decreased the contrast of the dust features in
these color maps.  This resolution mismatch, however, introduces small
uncertainties into our estimate of the amount of reddening present in
the color maps. We discuss this further in section \S 4.3 below.

The images shown in Figures 1, 2 \& 3 reveal a wide range of morphologies,
including nuclear spiral patterns, stellar bars, and irregular, clumpy
dust distributions.  In Table 2 we list our morphological classification for
each galactic nucleus. The most
common morphology we find in these images are nuclear spiral dust lanes,
generally extending into the inner $10 - 100$ pc.  These nuclear spiral
arms are wound in the same direction as the main disk spiral arms,
but they do not appear to be continuations of the main disk spiral arms.
In many of the galaxies with large-scale bars, these nuclear dust spirals are
clearly connected to dust lanes along the leading edge of the bar, similar to
what is seen in the hydrodynamic simulations of Athanassoula (1992). The most
striking examples are Mrk\,471, UGC\,12138, NGC\,5674, NGC\,3362, NGC\,5347,
and NGC\,7674.

The nuclear spiral dust lanes in Figures 1, 2, \& 3 exhibit a variety of
spiral morphologies, ranging from relatively
tightly wound arms (e.g., NGC\,1144 and UGC\,6100) to loosely wound arms
(e.g., UGC\,12138 and Mrk\,573).  We find examples of one-sided spiral arms
(e.g., NGC\,3982 and Mrk\,270), two-arm,  ``grand-design'' spirals
(e.g., UGC\,12138 and NGC\,7682), and multi-armed, ``flocculent'' spirals
(e.g., Mrk\,334 and NGC\,5674).  Often it is easier to see spiral arms on one
half of the image than the other, indicating that the gas disks containing the
spiral structure are inclined to the line of sight and we are more clearly
seeing the dust lanes on the near side of the galaxy (e.g. Mrk\,744 and
NGC\,1144).

In some of our color maps we also see regions with very blue colors.  These
include diffuse ``blue'' regions that are due to extended emission-line gas
and bright, unresolved knots of blue light that are likely regions of OB star
formation.  In particular, these latter regions are found in
Mrk\,334, Mrk\,744, Mrk\,266SW, and NGC\,7682 (Figures 1 \& 2).
While we do not have UV imaging to confirm that there is significant star
formation in these sources, some authors
(Heckman et al. 1995, 1997; Gonz\'alez Delgado et al. 1998) have proposed
that massive star formation on 100 pc scales may be a significant contributor
to the continuum emission in some Seyfert 2s. These galaxies may therefore be
candidates for such Seyfert 2--Starburst galaxies.

\subsection{Nuclear Bars}

Nuclear stellar bars have been implicated as a possible mechanism for
removing angular momentum from gas on 100's of parsec scales and
transporting it to the inner 10's of parsecs, where the central
supermassive black hole begins to dominate the gravitational potential.
Recent ground-based (Mulchaey \& Regan 1997) and {\it HST} observations (Regan
\& Mulchaey 1999) have shown that most Seyfert galaxies do not have nuclear
bars, implying that some other mechanism is responsible for transporting
interstellar gas to the central engine.  Our data confirm and extend their
conclusions with a larger, representative sample of Seyfert 2s.
We searched for nuclear bars by qualitatively examining the isophotal contour
maps of our $V$, $J$ and $H$ images. In several galaxies we also found
straight dust lanes in the color maps, but in all cases the bar signature
was clearly visible in the isophotal contour maps (see Figure 4).  In a future
paper we will present a more quantitative analysis of bar selection in
these galaxies.

We find at most 5 nuclear bars among the 24 galaxies in our sample (see
Figure 4).  Mrk\,573 is the best example of a double-barred galaxy in our
sample. It has both a host-galaxy bar and a nuclear bar as seen in previous
ground-based and {\it HST} visible-wavelength imaging (Pogge \& DeRobertis 1993;
Capetti et al. 1996), and there is a clear straight dust lane going into
the nucleus in visible and visible/near-IR color maps (Pogge 1996;
Capetti et al. 1996; Quillen et al. 1999).  The appearance of this
dust lane in the $V-H$ color map is consistent with dust lanes
at the leading edge of the bar (cf. Quillen et al. 1999).  The nuclear bar
in Mrk\,270 is apparent at visible and infrared wavelengths, and
there is a straight dust lane in the $V-H$ color map oriented
nearly perpendicular to the bar major axis. NGC\,5929 has a distinct nuclear
bar in the near-infrared images, but it is hidden by dust in the
visible-light images.  The $V-H$ color map of this galaxy shows
the nuclear region is very dusty with an irregular dust morphology.
Mrk\,471 shows evidence for a nuclear bar in both the near-infrared
surface brightness image and through straight dust lanes in the $V-H$ color
map. The lowest surface brightness contours in Figure 4 also show the
large-scale bar in Mrk\,471.

Regan \& Mulchaey (1999) report a nuclear bar in NGC\,5347.  While this
galaxy has a clear large-scale bar seen at visible and IR wavelengths
(McLeod \& Rieke 1995), the nuclear bar is less obvious in the $H-$band image
(also shown in Figure 4).  While we will retain the
classification of this galaxy as nuclear barred in the interests of setting
a strong upper limit on the nuclear bar fraction in Seyferts, we consider
this the weakest case for a nuclear bar of the 5 we have found.

Peletier et al. (1999) have suggested that there is a nuclear bar
perpendicular to the main galaxy bar in NGC\,5674 at visible wavelengths,
but we do not see it in our near-infrared images (Figure 4).  Instead, we
find a great
deal of dust in this galaxy at the position angle of both semiminor axes
and thus the ``bar'' apparent at visible wavelengths is most likely an
artifact of the circumnuclear dust distribution.  This underscores the
advantage of using near-infrared images to search for and confirm suspected
nuclear bars, as dust can greatly distort the nuclear surface brightness
profiles at visible wavelengths.

Finally, nuclear bars that are heavily enshrouded by dust may not reveal
their presence with straight dust lanes in $V-H$ color maps. If nuclear
bars are generally hidden by large quantities of dust, however, straight
dust lanes may be visible in $J-H$ color maps. We have created $J-H$ color
maps for all of the galaxies in our sample, but find no additional evidence
for nuclear bars in the $J-H$ color maps that were not detected in
our $V$, $H$, or $V-H$ images.

\subsection{Circumnuclear Dust Lanes}

Seyferts, like more quiescent spirals, have a central stellar component that
consists of a disk and a central bulge.  In the nuclear regions there is
additional light from the AGN and the extended narrow-line region.  This latter
region is partially resolved in many {\it HST} images of Seyferts (Bower et al.
1994; Simpson et al. 1997; Malkan et al. 1998).  The near-infrared continuum
emission from the unresolved active nucleus proper may also include a
contribution from hot dust (e.g., Glass \& Moorwood 1985; Alonso-Herrero et al.
1998; Glass 1998).

The presence of circumnuclear dust manifests itself in our
images in two ways.  First, we expect there to be a relatively uniform
distribution of diffuse dust spread throughout the volume of the nuclear
bulge and disk of the host galaxy.  This component will be
difficult to measure as it is expected to be well mixed with the stars.
Second, we expect discrete clouds of gas and dust, most likely organized
into a disk confined to a plane (or warped slightly), which could give rise
to the spiral dust lanes seen in the color maps (Figures 1, 2, \& 3).

Diffuse gas and dust uniformly distributed in the central regions of the
galaxies would have a
relatively low integrated mass, and so is unlikely to be a significant
contributor to the fuel reserves of the active nuclei.  The discrete, dense
dust structures we see in the nuclear disks, however, particularly in the
spiral dust lanes, are a more probable fuel source.  As these spiral dust lanes
are likely to be confined to a disk, their net extinction can be estimated
by treating them as a thin, obscuring slab of material and using the
change in $V-H$ color between the arm and interarm regions as an estimate
of the total extinction, $A_V$.
We define the
color excess due to dust in the nuclear disk as: $$E(V-H) = (V-H)_{arm} -
(V-H)_{interarm}.$$ This excess color is used to estimate the total
visual extinction, $A_V$, via a standard interstellar extinction curve
(Mathis 1990). The $J-H$ color excess could also be used for this purpose, but
due to the close proximity of these filters in wavelength, they do not provide
as precise a measure of $A_V$.  By relating this $A_V$ to a column density of
hydrogen atoms, $N_H = 1.87 \times 10^{21}$ cm$^{-2} A_V$ for an $R_V = 3.1$
reddening law (Bohlin, Savage,  \& Drake 1978) and converting this into a mass
column density, we can obtain a rough estimate of the surface density of
these disks: $$\Sigma = 15 \times A_V\;M_{\sun}\;{\rm pc}^{-2}.$$
The disk surface density derived by this technique actually estimates the
maximum surface density rather than the average.
However, as discussed in
detail below, this measure of $E(V-H)$ will tend to underestimate the true
value of $A_V$ due to uncertainties in the dust geometry and instrumental
effects.  We therefore chose this conservative approach to estimating the
average disk surface density so as to minimize the likelihood that we have
significantly underestimated the surface density.

While this estimate is computationally straightforward, there are several
systematic effects at work that combine to make these estimates lower limits
on the true gas surface density.  First, we have used a very simple foreground
screen as our dustlane model.  A more physically realistic model would be a
dust screen sandwiched between two layers of stars, corresponding to the stars
behind and in front of the dust relative to the line of sight.  For an equal
amount of starlight on either side of the dust lane, the arm-interarm
$E(V-H)$ defined above reaches a maximum of $E(V-H) \sim 0.45$ at
$A_V \sim 1.5$ for a smooth dust layer.  A patchy or clumpy distribution,
as discussed below, would increase these numbers.  Scattering is another
vital component to a realistic dust model.
Dust particles are strongly forward scattering and more
likely to scatter blue light than red light, causing a ``bluing'' of the
starlight that could counteract, in part, the reddening due to absorption.
Significant scattering would reduce the observed $E(V-H)$, thus
leading us to underestimate the total column density of material in
these nuclear spiral arms.  We estimated the magnitude of scattering by
incorporating the ``Dusty Galactic Nuclei'' model of Witt, Thronson, \& Capuano
(1992) into our dust sandwich model.  We find that scattering can decrease
the observed $E(V-H)$ by a factor of two, corresponding to an increase in the
disk surface density by the same factor.  The simple foreground screen
model we have used to estimate the disk surface density therefore results in
an underestimate of the true disk surface density by a factor which depends
on the true dust geometry, but is useful as it sets a lower limit on the
surface density.

Another complication to our estimate of the disk surface density is that
our images do not resolve
small-scale dust structures in these dust lanes. Studies of the
arm-interarm extinction in overlapping galaxies (White \& Keel 1992; White,
Keel, \& Conselice 1996; Berlind et al. 1997) show that while the large-scale
dust lanes are optically thin, the extinction law in the spiral arms
tends to be grayer than a standard Galactic extinction law, primarily due
to unresolved, optically thick clumps (Berlind et al. 1997). Unresolved
clumps of dust in these nuclear spirals, particularly in the more distant
galaxies, would lead to a grayer extinction law than assumed here,
resulting in an underestimate of the total dust mass.

A second resolution effect is due to the finite resolution of the WFPC2 and
NICMOS cameras.  Many of the dust lanes in our $V-H$ color maps are unresolved
or only marginally resolved. We have therefore constructed a simple model of
a dust lane to estimate the magnitude of this effect. We constructed models
of dust lanes by creating
a series of artificial images at $V$ and $H$ of ``ideal'' dust lanes in which
the dust lane was a step function in flux at a range of widths and
$E(V-H)$ (note that $A_V = 1.2 E(V-H)$ for $R_V = 3.1$; Mathis 1990). We then
convolved these images with the {\it HST} PSF for the PC1 camera and NICMOS
Camera 1 from TinyTim
(Krist \& Hook 1999) and measured the difference
$E(V-H)_{model} - E(V-H)_{conv}$ as a function of the width of the unconvolved
dust lane in units of PC1 pixels. In Figure 5 we plot the difference
in $E(V-H)$ to illustrate the effect of unresolved and marginally resolved
dust lanes on measurements of the extinction ({\it solid lines}). The lines
correspond to input $E(V-H)$ values of 1.1, 0.8, and 0.4 with the top line
corresponding to the largest $E(V-H)$ we measured.  This figure shows that
unresolved dust lanes may cause us to underestimate $A_V$ by up to
a factor of three for the most heavily reddened dust arms in this sample.

A further complication is the resolution mismatch between the PC1 and NICMOS
Camera 1, which corresponds different widths for the $V$ and $H$ PSFs.
Creating a $V-H$ color map corresponds to dividing a $V-$band image of the
galaxy that has been convolved with the narrower WFPC2 PSF by an $H-$band
image that has been convolved with the broader NICMOS PSF. The result is a
weak unsharp masking effect that causes the $V-H$ color to be slightly larger
in unresolved or marginally resolved dust lanes.  To measure the size of
this effect, we repeated the exercise discussed above, but convolved both the
$V$ and $H$ band image with the same PSF (in this case the PSF for the F606W
filter and PC1 detector).  We have plotted the results of
this test ({\it dotted lines}) on Figure 5 for the same $E(V-H)$ values and, as
expected, the model dust lanes convolved with different PSFs are systematically
redder than the model dust lanes convolved with identical PSFs.  The magnitude
of this effect, however, is much less than that due to convolving the model
dust lanes with a PSF in the first place.

A final factor that introduces systematic uncertainties into the measurement of
$E(V-H)$ is the possible presence of diffuse line emission. As mentioned
above, we are resolving the narrow-line region in several of these
galaxies. In particular, Mrk\,270, Mrk\,573, NGC\,3362, NGC\,7682, and
UGC\,6100 all show bright, extended line emission in the F606W images that
result in ``blue'', and often cone-shaped, structures in their $V-H$ color
maps.  The bow-shaped emission line region in Mrk\,573 and other
Seyfert 2s have been previously studied with {\it HST} by Falcke, Wilson \&
Simpson (1998). Capetti et al. (1996) have seen similar structures in {\it HST}
images of the narrow line regions in Seyfert 2s.
In addition to these discrete structures, there may also be faint, diffuse
line-emitting gas spread throughout the circumnuclear regions. Since the F606W
filter contains the bright [O III] $\lambda 5007$ and H$\alpha$ lines,
a significant emission-line contamination is possible. We estimated the
contribution of
emission lines to a variety of Seyfert 2 spectra with the STSDAS SYNPHOT
package. This exercise showed that if these lines have a total equivalent width
of 100 \AA, they would still only increase the $V-$band surface brightness
by $\Delta~V = -0.07$ magnitudes. The contribution to the total flux in the
F606W filter is small
because the filter has an effective width of approximately 1500 \AA.
The brightest contributors to the $H-$band surface brightness, the hydrogen
Brackett lines, are significantly lower equivalent width than their
visible-wavelength counterparts and therefore the emission-line
contribution to the $H-$band is expected to be unimportant. In addition to
requiring a very
high equivalent width to affect the $V-$band flux, this emission would
have to be correlated or anticorrelated with the dust lanes to lower
or raise, respectively, the measured $E(V-H)$.

To summarize our results, our ($V-H$) color maps of these galaxies exhibit
dust lanes with a range in $E(V-H)$ corresponding to inferred mass surface
densities of $1-20\; M_{\sun}$~pc$^{-2}$.  While we have overestimated the
average surface density by measuring the maximum value of $E(V-H)$,
the dust model and resolution effects could offset this as they lead to an
underestimate of the total column density.
These surface densities imply that
these nuclear disks contain at least $10^6\;M_{\sun}$ of interstellar material
in the inner $100 - 200$ pc.  As Seyfert galaxies require mass accretion rates
of $\sim 0.01 M_{\sun}$ yr$^{-1}$ (e.g. Peterson 1997) to fuel their nuclear
luminosities, these circumnuclear disks are massive enough to act as potential
reservoirs for fueling the AGN.

\section{Nuclear Spirals and Fueling}

Nuclear spirals like those we have shown here have been observed in
quiescent spirals (Phillips et al. 1996; Carollo, Stiavelli, \& Mack 1998;
Elmegreen et al. 1998; Laine et al. 1999), as well as in Seyfert galaxies
(Quillen et al. 1999;
Regan \& Mulchaey 1999).  These nuclear spirals are dynamically distinct
from the large-scale spiral arms seen in spiral galaxies as they may lie
inside the inner Lindblad Resonance of the outer spiral arms (Elmegreen,
Elmegreen, \& Montenegro 1992) and they may also be shielded by the
dynamical influence of the galaxy's nuclear bulge (Bertin et al. 1989).
The nuclear spiral dust lanes appear to differ from the large-scale
spiral structure of their hosts in two important ways: they are unlikely to
be the result of gravitational instabilities in self-gravitating disks, and
their sound speed is likely to be comparable to the orbital speed.

Elmegreen et al. (1998) estimated the value of Toomre's (1981) $Q$ parameter
for the nuclear disk in NGC\,2207 and determined that the nuclear disk was not
self-gravitating. We have applied a similar analysis to the galaxies in our
sample after rewriting the expression for $Q$ in the form:
$$Q = 1.5 \times \left(\frac{a}{10\;{\rm km}\;{\rm s}^{-1}}\right)^2
\left(\frac{\Sigma}{M_{\sun}\;{\rm pc}^{-2}}\right)^{-1}
\left(\frac{h}{{\rm pc}}\right)^{-1}$$
where $a$ is the sound speed, $\Sigma$ is the
azimuthally-averaged disk surface density in $M_{\sun}$ pc$^{-2}$, and $h$ is
the disk scale height in pc. This expression for $Q$ assumes that the nuclear
disks are in solid--body rotation, as is the case in NGC\,2207 (Rubin \& Ford
1983) and in the circumnuclear gas disks in galaxies with rapidly rotating
cores in the Virgo Cluster (Rubin et al. 1997).  We have estimated $a$ to be
10 km s$^{-1}$, which is
a reasonable value for the inner disks of spirals (e.g. Spitzer 1978;
Elmegreen et al. 1998), although some authors report values of
$3 - 10$ km s$^{-1}$ for the main disks of spirals (e.g. Kennicutt 1989;
van der Kruit \& Shostak 1984).
We are able to place an upper limit on $h$ by assuming that as the disks
are thinner than the width of the spiral arms, as is true
for spiral arms in galaxy disks proper.  Most of our estimates of the arm
widths are also upper limits, as at their narrowest points nearly all of the
nuclear spiral arms are unresolved.  Our estimates of $h, \Sigma$, and $Q$ are
summarized in Table 3.

Values of $Q<1$ would imply that the spiral arms are formed by gravitational
instabilities in self-gravitating disks (e.g. Binney \& Tremaine 1987),
whereas values of $Q>1$ would correspond to spiral arms formed in
non-selfgravitating disks by hydrodynamic instabilities.  The $Q$ values we
find for the spiral arms in our Seyfert galaxies range from $5-200$ (Table 3,
column 4).  We thus conclude, as did Elmegreen et al. (1998) for NGC\,2207,
that these nuclear spiral arms do not occur in self-gravitating gas disks.
Since $Q$ scales as the square of the
sound speed, a lower value of $a$ could greatly reduce $Q$.  As most
of our estimates for $Q$ are significantly above 1, $a$ would have to be
quite small to change our results in the direction of self-gravitating $Q$
values.  Faster sound speeds, in contrast, would increase $Q$ as $a^2$.
Our estimates of $\Sigma$ are rough values that could vary due to several
competing effects.  As mentioned above, we may be underestimating the column
density in these dust lanes due to scattering and the effects of unresolved
dust lanes.  In contrast, by measuring $E(V-H)$ from the maximum arm-interarm
$V-H$ contrast, we may be overestimating the disk surface density.
In addition, if the spiral arms are shock fronts they may correspond to density
enhancements of up to a factor of 4 (for nonradiating, adiabatic shocks) or
higher (for radiating shocks; e.g. Spitzer 1978) over the azimuthally-averaged
disk surface density.  In addition to the uncertainty in $a$ and $\Sigma$,
our estimate of the scale height is an upper limit as the spiral arms are
generally unresolved.  However, decreasing $h$ would increase $Q$ and the
upper limit on $h$ thus only strengthens the argument that these disks are
not self-gravitating.  The fact that
many of these nuclear spirals are multi-arm spirals further supports the
notion that they are unlikely to be due to gravitational instabilities in
self-gravitating disks (Lin \& Shu 1964).

Spiral arms can form in gaseous disks that are not self-gravitating through
hydrodynamic shocks.  One possibility is that these shocks could be caused by
turbulent gas motions driven by the inflow of material from larger radii
(Struck 1997).
Englmaier \& Shlosmann (1999) have also shown that shocks in
non-selfgravitating, nuclear disks can form grand-design, two-arm spiral
structure in galaxies with a large-scale stellar bar.
Since the orbital speeds in the inner regions of galaxies are of order the
sound speed, it is also possible that the arms are formed by acoustic shocks
(Elmegreen et al. 1998; Montenegro, Yuan, \& Elmegreen 1999). Shocks in these
gaseous disks could be a viable means of removing angular momentum and energy
from the gas and thus these nuclear spirals could be signposts of the fueling
mechanisms for Seyfert galaxies.
While investigations have traditionally looked for large-scale
signatures of interaction or bars in the host galaxies, the actual funneling
of gas the last few hundred parsecs into the central black hole may be mediated
by the hydrodynamics of non-selfgravitating gas disks on small scales.  This
mechanism could be common, regardless of whether large-scale bars or
interactions transported the gas from large radii to 100's of parsec scales.

\section{Summary and Conclusions}

These $H$-band images and $V-H$ color maps of the centers of a representative
sample of 24 Seyfert 2 galaxies rule out the presence of nuclear bars in
all but at most 5 of the 24 galaxies in our sample.  While this
is not to say that nuclear bars cannot play some role in fueling an AGN if
present, they are found in only a minority of Seyferts.  This strengthens
and extends similar results found with a smaller sample of Seyferts by
Regan \& Mulchaey (1999). While this sample is composed of only Seyfert 2s,
a study of all of the Seyfert 1s and 2s in the CfA Seyfert sample with
{\it HST} WFPC2 imaging did not reveal an excess of bars among Seyfert 1s
relative to Seyfert 2s (Pogge \& Martini 2000).  We note, however, that the
circumnuclear environments of Seyfert 1s are more difficult to study given
their strong nuclear PSFs and they have not been imaged in the
near-infrared, which was needed to find 2 of the 5 possible bars in our
sample.

Our $V-H$ color maps of the centers of these Seyfert galaxies have shown
that 20 of 24 have nuclear spiral structure. Of the 4 that do not show
nuclear spiral structure, 2 of these (NGC\,4388 and NGC\,5033) are nearly
edge-on systems and thus the geometry is unfavorable for viewing any kind
of nuclear spiral structure.  The remaining two, Mrk\,266SW and NGC\,5929,
are both in strongly interacting systems and have very chaotic dust
morphologies, perhaps reflecting disordered delivery of host galaxy gas due to
the tidal interaction.  We have used the $V-H$ color maps to estimate the
amount of material in these spiral arms and have found that the nuclear
disks in which they reside contain enough gas to serve as a fuel
reservoir for the AGN. The surface densities of these disks, along with our
estimates of the disk scale height and surface density, also suggest that
these disks are not self-gravitating. The nuclear spiral structure is
therefore probably formed by shocks propagating in the disk, via either an
acoustic mechanism or other hydrodynamic process, and these shocks could
dissipate energy and angular momentum from the gas.

Our observations strongly rule out small-scale nuclear bars as the primary
means of removing angular momentum from interstellar gas to fuel the AGN in
our representative sample of Seyfert 2s.  Instead, we find that the most common
morphological features in the centers of AGN are nuclear spirals of dust.
These spirals may be caused by shock waves in a disk or may be
streamers of mass falling inwards from the innermost orbital resonance. The
large fraction of Seyfert 2s with dusty nuclear spirals demonstrates the
importance of shocks and gas dynamical effects in removing angular momentum
from the gas and feeding this material into the nucleus and hence into
the supermassive black hole.

An outstanding question posed by these nearly ubiquitous nuclear spirals
is how their host nuclear disks formed.  As proposed by Rubin et al. (1997)
for nuclear disks in some Virgo Cluster galaxies, we conclude that two likely
scenarios are the standard processes previously invoked for fueling AGN:
bars and interactions.  Both large-scale bars and interactions can remove
angular momentum from gas and drive it into the inner 100's of parsecs, where
it settles into a nuclear disk.  Shocks or other hydrodynamic effects in these
disks then complement the action of bars and interactions by removing angular
momentum from this accumulated material and feeding the fuel into the active
nucleus.

Two future areas of work are suggested by our results. First, similar
observations of a large sample of quiescent spiral galaxies should be
obtained to determine if the nuclear spirals found in Seyferts are present in
all spirals.  This work should also be augmented by spectroscopic data
to provide kinematic information on the gas and dust in the inner
regions. Second, theoretical predictions of accretion rates from spiral
structure in both self-gravitating and non-selfgravitating disks are
needed. Answers to these questions are needed before we can understand the role
played by dusty nuclear spirals in fueling AGN.

\acknowledgments

We would like to thank Alice Quillen for many useful discussions and comments
and Brad Peterson and Barbara Ryden for their helpful comments on the
manuscript.  We would also like to thank the referee, Mike Regan, for
his helpful comments that have improved and clarified this paper.
Support for this work was provided by NASA through grant numbers
GO-07867.01-96A and AR-06380.01-95A from the Space Telescope Science
Institute, which is operated by the Association of Universities for Research
in Astronomy, Inc., under NASA contract NAS5-26555.
This research has made use of the
NASA/IPAC Extragalactic Database (NED) which is operated by the Jet Propulsion
Laboratory, California Institute of Technology, under contract with the
National Aeronautics and Space Administration.

{}

\begin{deluxetable}{lccccccc}
\footnotesize
\tablenum{1}
\tablewidth{500pt}
\tablecaption{Sample Characteristics}
\tablehead {
  \colhead{Name} &
  \colhead{Alternate} &
  \colhead{Seyfert Type} &
  \colhead{Morphological Type} &
  \colhead{Note} &
  \colhead{WFPC2 Filter} &
  \colhead{Distance} &
  \colhead{pc/$''$} \\
}
\startdata
Mrk 334  &       & 1.8   & Pec          & Disturbed     & F606W &  91.5 &444\nl
Mrk 471  &       & 1.8   & SBa          &               & F606W & 136.9 &664\nl
Mrk 744  &NGC 3786& 1.8  & SAB(rs)a pec & Interacting   & F606W &  36.1 &175\nl
UGC 12138&2237+07  & 1.8 & SBa          &               & F606W & 102.8 &498\nl
NGC 5033 &       & 1.9   & SA(s)c       & Edge On       & F547M &  12.1 &59\nl
NGC 5252 &       & 1.9   & SO           &               & F606W &  90.7 &440\nl
NGC 5273 &       & 1.9   & SA(s)0$^0$   &               & F547M &  14.9 &72\nl
NGC 5674 &       & 1.9   & SABc         &               & F606W &  98.1 &476\nl
UM 146   &0152+06& 1.9   & SA(rs)b      &               & F606W &  71.6 &347\nl
Mrk 461  &       & 2     &              &               & none  &  65.6 &318\nl
Mrk 266  &NGC5256& 2     & Comp Pec     &               & F606W & 111.0 &538\nl
Mrk 270  &NGC5283& 2     & SO?          &               & F606W &  38.2 &185\nl
Mrk 573  &       & 2     & (R)SAB(rs)O+:&               & F606W &  71.0 &344\nl
NGC 1068 &       & 2     &              &               & F606W &  16.2 &79\nl
NGC 1144 &       & 2     & RingB        &               & F606W & 116.5 &565\nl
NGC 3362 &       & 2     & SABc         &               & F606W & 108.8 &527\nl
NGC 3982 &       & 2     & SAB(r)b:     &               & F606W &  17.1 &83\nl
NGC 4388 &       & 2     & SA(s)b: sp   & Edge On       & none  &  32.2 &156\nl
NGC 5347 &       & 2     & (R')SB(rs)ab &               & F606W &  31.4 &152\nl
NGC 5695 &Mrk 686& 2     & SBb          &               & F606W &  56.9 &276\nl
NGC 5929 &       & 2     & Sab: pec     & Interacting   & F606W &  34.9 &169\nl
NGC 7674 &Mrk 533& 2     & SA(r)bc pec  &               & F606W & 118.5 &575\nl
NGC 7682 &       & 2     & SB(r)ab      &               & F606W &  70.8 &343\nl
UGC 6100 &A1058+45& 2    & Sa?          &               & F606W & 117.6 &570\nl
\enddata
\tablecomments{
Properties of the galaxies observed with the NICMOS Camera 1.
Columns 1 \& 2 list the most common names for the targets, and column 3
lists its Seyfert type as defined by Osterbrock \& Martel (1993). In
column 4 we have compiled the morphological type for the host galaxy
from NED, while in column 5 we have provided additional morphological
information. Column 6 lists the visible-band filter of the archival image used
in the color map. Column 7 lists the distance of the galaxy in Mpc
(assuming $H_0 = 75$ km s$^{-1}$ Mpc$^{-1}$) and column 8 gives the projected
size in pc of $1''$ at the distance of the galaxy.
}
\end{deluxetable}
\normalsize

\footnotesize
\begin{deluxetable}{lll}
\footnotesize
\tablenum{2}
\tablewidth{500pt}
\tablecaption{Nuclear Taxonomy}
\tablehead {
  \colhead{Name} &
  \colhead{Bar Type} &
  \colhead{Nuclear Morphology} \\
}
\startdata
Mrk 334  & None & Nuclear Spiral \nl
Mrk 471  & Nuclear & Nuclear Spiral  \nl
Mrk 744  & Host & Nuclear Spiral \nl
UGC 12138& Host & Nuclear Spiral \nl
NGC 5033 & None & Edge On    \nl
NGC 5252 & None & Nuclear Spiral \nl
NGC 5273 & None & Nuclear Spiral  \nl
NGC 5674 & Both & Nuclear Spiral \nl
UM 146   & None & Nuclear Spiral \nl
Mrk 461  & None & No Visible \nl
Mrk 266  & None & Irregular   \nl
Mrk 270  & Nuclear & Nuclear Spiral \nl
Mrk 573  & Both & Nuclear Spiral \nl
NGC 1068 & Host & Nuclear Spiral \nl
NGC 1144 & Ring & Nuclear Spiral \nl
NGC 3362 & Host & Nuclear Spiral \nl
NGC 3982 & Host & Nuclear Spiral \nl
NGC 4388 & None & Edge On \nl
NGC 5347 & Host & Nuclear Spiral \nl
NGC 5695 & Host & Nuclear Spiral \nl
NGC 5929 & Nuclear & Nuclear Spiral \nl
NGC 7674 & None & Nuclear Spiral \nl
NGC 7682 & Host & Nuclear Spiral \nl
UGC 6100 & None & Nuclear Spiral \nl
\enddata
\tablecomments{
This table lists the nuclear properties of the galaxies in the sample based
upon our classification of the $V-H$ color maps. Column 1 lists the name of
the galaxy as in Table 1, while column 2 lists whether a host galaxy and/or
nuclear bar is present. Column 3 lists our morphological classification for
the galaxy based upon the dust structure shown in our color maps.
}
\end{deluxetable}
\normalsize

\footnotesize
\begin{deluxetable}{lccc}
\footnotesize
\tablenum{3}
\tablewidth{500pt}
\tablecaption{Nuclear Disk Parameters}
\tablehead {
  \colhead{Name} &
  \colhead{$\Sigma$ ($M_{\sun}$ pc$^{-2}$)} &
  \colhead{$h$ (pc)} &
  \colhead{$Q$} \\
}
\startdata
Mrk 334  & 20 & 141 &   5 \nl
Mrk 471  & 16 &  60 &  15 \nl
Mrk 744  & 19 &  16 &  48 \nl
UGC 12138& 14 & 113 &  10 \nl
NGC 5252 &  9 &  40 &  41 \nl
NGC 5273 & 17 &   7 & 130 \nl
NGC 5674 & 15 &  87 &  11 \nl
UM 146   & 18 &  32 &  26 \nl
Mrk 270  &  6 &  17 & 137 \nl
Mrk 573  &  9 &  63 &  26 \nl
NGC 1144 & 15 &  51 &  19 \nl
NGC 3362 &  9 &  48 &  34 \nl
NGC 3982 &  9 &   8 & 215 \nl
NGC 5347 & 18 &  42 &  20 \nl
NGC 5695 & 14 &  25 &  43 \nl
NGC 7674 & 12 & 105 &  12 \nl
NGC 7682 &  9 &  31 &  52 \nl
UGC 6100 & 12 &  52 &  24 \nl
\enddata
\tablecomments{The nuclear disk parameters for all galaxies showing nuclear
spiral structure. Column 1 contains the name of the galaxy, as listed in
tables 1 and 2. Column 2 has our estimate of the mass surface density based
on the $E(V-H)$ color, which is a lower limit. Column 3 is an upper limit to
the scale height of the disk based on measuring the width of the spiral
arms. Both of these quantities have been derived assuming the distance
listed in Table 1. Column 4 lists the value for the Toomre $Q$ parameter
we derived for each disk as discussed in section 6.
}
\end{deluxetable}
\normalsize

\clearpage

\begin{figure}
\plotone{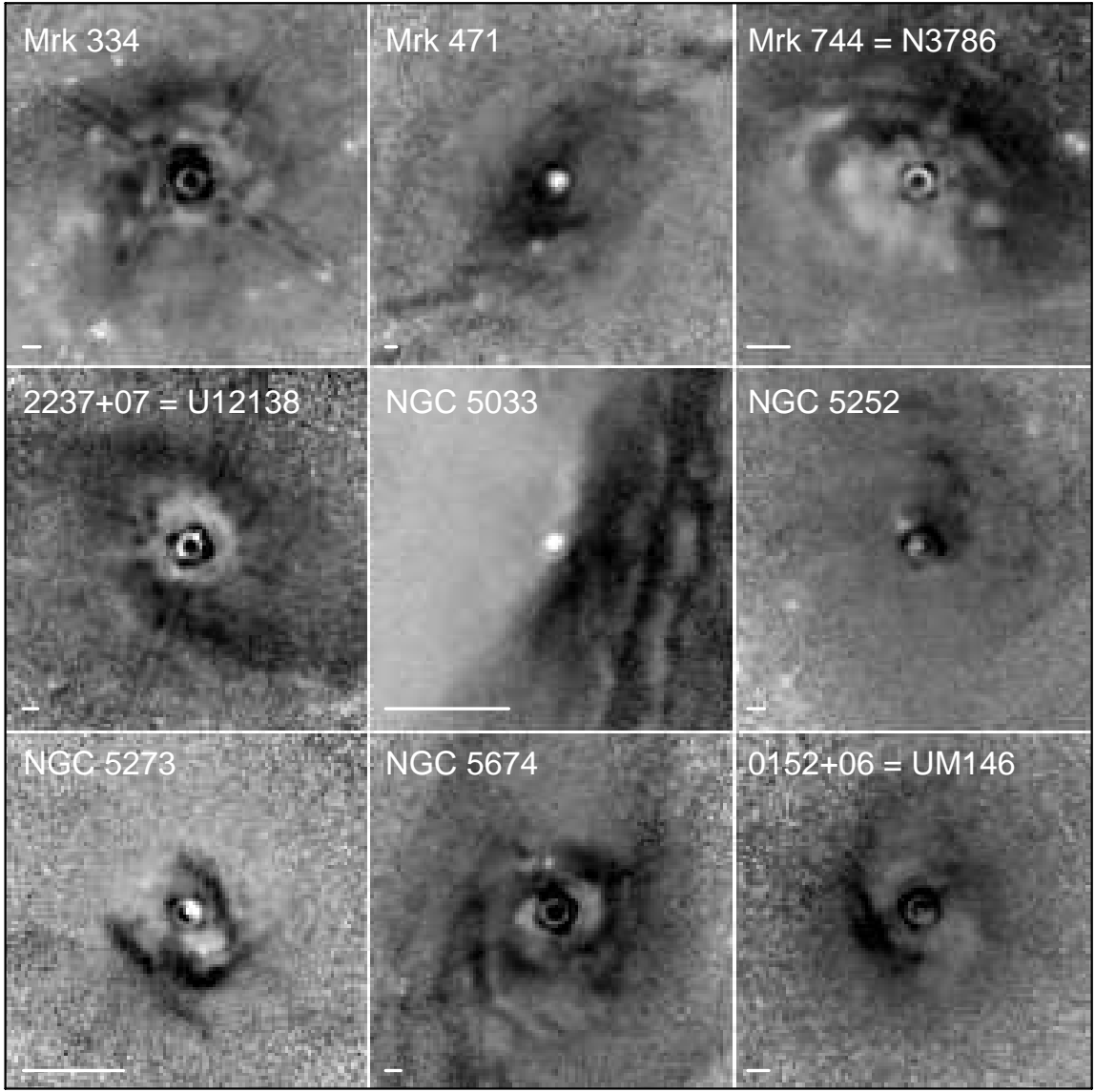}
\caption{$V-H$ color maps of the all of the Seyfert 1.8s and 1.9s from our 
sample. Nuclear spiral structure is clearly seen in most of the images at this 
gray scale, where white corresponds to a color of $V-H \approx 2$ mag and black 
corresponds to $V-H \approx 6$. 
All of these images have been rotated so that north is up and 
east is to the left and all of the images are 5\arcsec\,on a side. The white 
bar in the lower left corner of each image corresponds to $100\;h_{75}$ pc 
at the distance of the galaxy.  In Table 2 we list the 
nuclear morphological classification for each of these objects. }
\end{figure}
 
\begin{figure}
\plotone{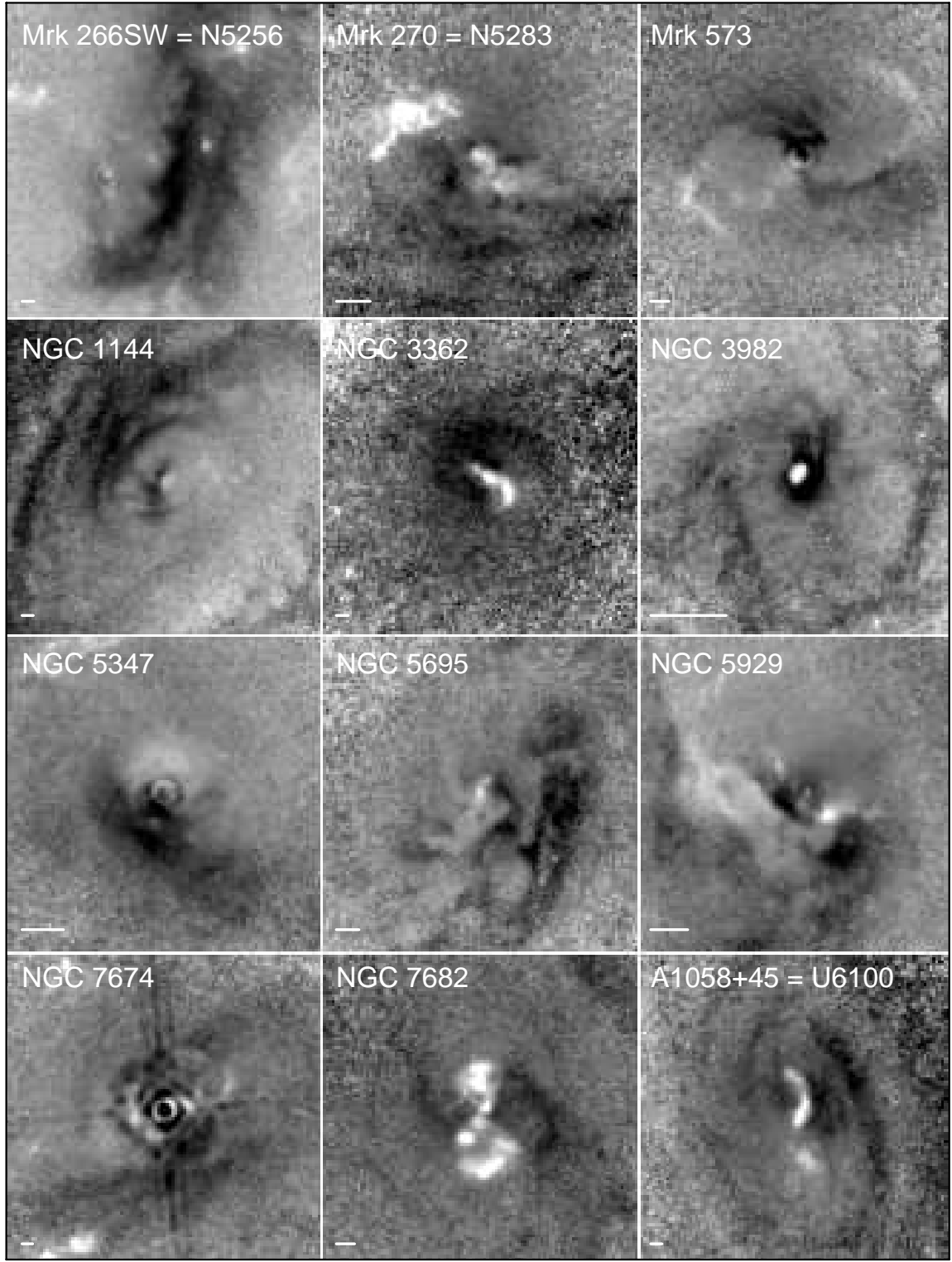}
\caption{Same as Figure 1, but for the Seyfert 2 galaxies. }
\end{figure}

\begin{figure}
\plotone{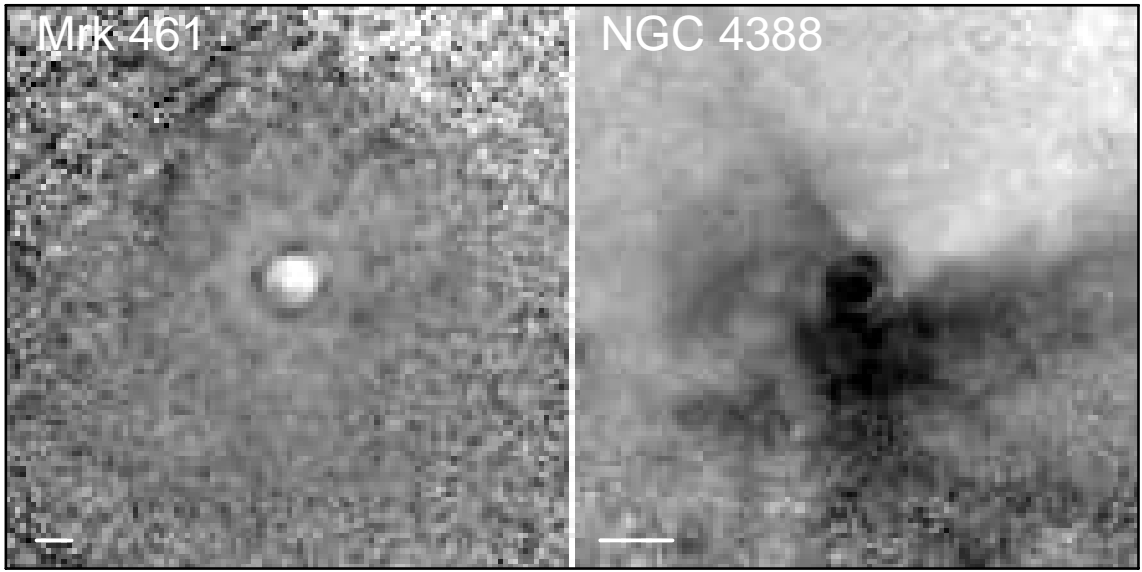}
\caption{$J-H$ color maps of the two galaxies in our sample without 
visible-band {\it HST} imaging. As in Figures 1 \& 2, these images have been 
rotated so that North is up and East is to the left. Both images are 5\arcsec\, 
on a side and the bar in the lower left corner corresponds to $100\;h_{75}$ pc
at the distance of the galaxy.}
\end{figure}

\begin{figure}
%\plotone{fig4.eps}
\plotfiddle{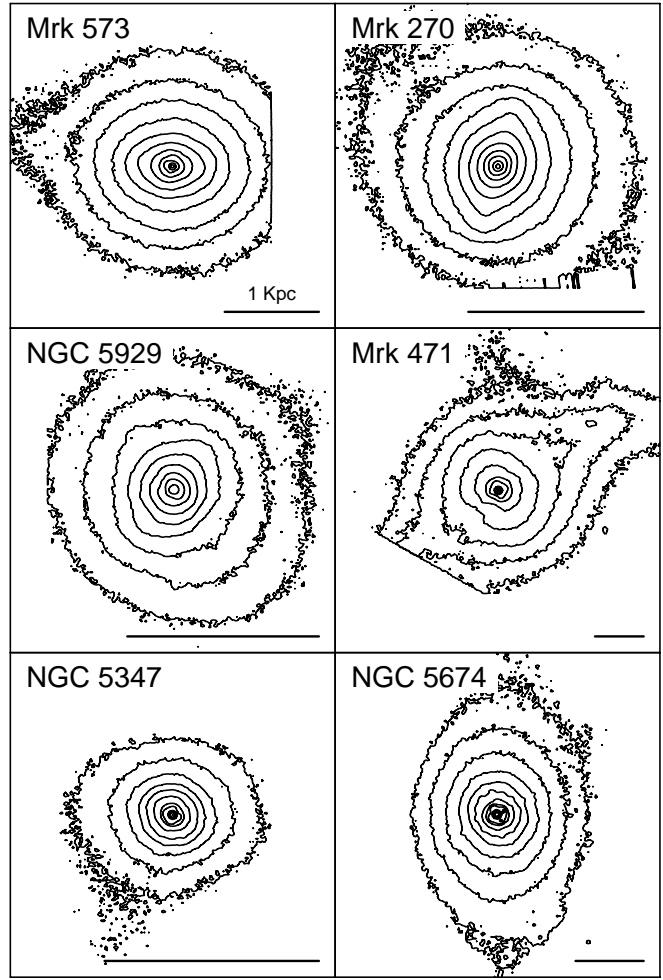}{5.0truein}{0}{60}{60}{-200}{0}
\caption{$H-$band contour maps of our bar canditates, rotated so that 
North is up and East is to the left. These images are 10\arcsec\, on a side and 
the bar at the bottom of the Figure corresponds to a projected distance of 
$1\;h_{75}$ kpc at the distance of the galaxy.  The abrupt edges in the 
contours of Mrk\,573, Mrk\,270, Mrk\,471 correspond to the edge of the NICMOS 
Camera 1 detector. }
\end{figure}

\begin{figure}
\plotone{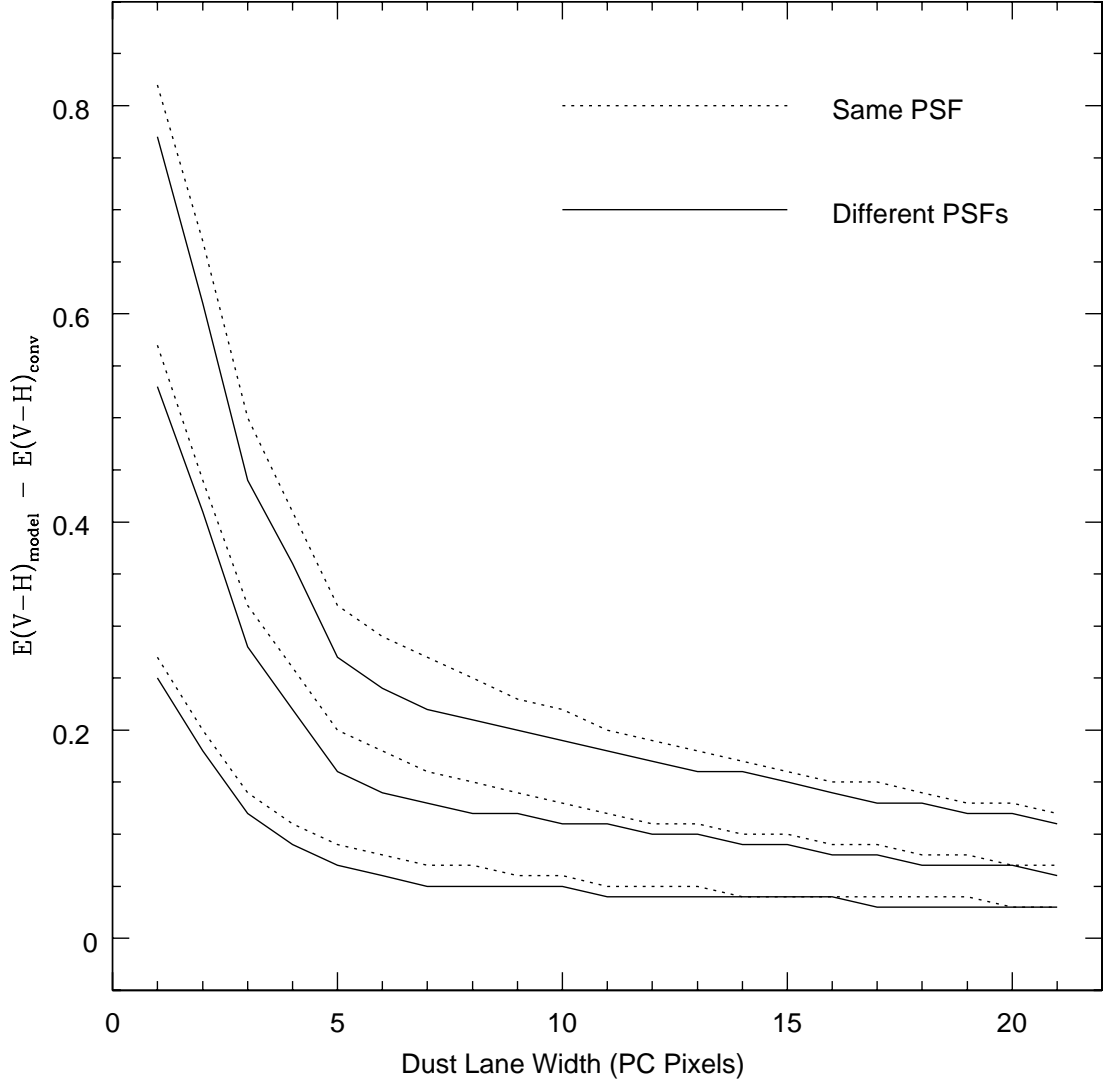}
\caption{The effect of convolving an unresolved dust lane by the {\it HST} 
PSFs for the PC1 and NICMOS Camera 1. The $x-$axis gives the width of the dust 
lane prior to convolution with the {\it HST} PSFs. The $y-$axis shows the 
difference between the input model $E(V-H)$ and the measured $E(V-H)$ after 
convolution. This shows that the convolution process causes the $E(V-H)$ to be 
significantly decreased by convolving it with the two PSFS ({\it solid lines}) 
for $E(V-H) = 1.1, 0.8,$ and $0.4$ mag from top to bottom. Convolving the dust 
lanes with the same PSF results in a slightly smaller correction ({\it dotted 
lines}). See section 4.3 for further details. 
}
\end{figure}

\end{document}